\documentclass[preprint]{aastex}

\shorttitle{Milky Way Project Yellowballs}
\shortauthors{Kerton et al.}

\begin{document}

\title{THE MILKY WAY PROJECT: WHAT ARE YELLOWBALLS?}

\author{C. R.  Kerton}
\affil{Department of Physics \& Astronomy, Iowa State University, 12  Physics Hall,  Ames, IA 50011, USA}
\email{kerton@iastate.edu}

\author{G. Wolf-Chase\altaffilmark{1}}
\affil{Astronomy Department, Adler Planetarium, 1300 S. Lake Shore  Drive,  Chicago, IL 60605, USA}
\email{gwolfchase@adlerplanetarium.org}

\author{K. Arvidsson}
\affil{Trull School of Science and Mathematics, Schreiner University,  2100 Memorial Blvd.,  Kerrville, TX 78028, USA}
\email{KDArvidsson@schreiner.edu}

\author{C. J. Lintott and R. J. Simpson}
\affil{Oxford Astrophysics, Denys Wilkinson Building, Keble Road,  Oxford OX1 3RH, UK}
\email{cjl@astro.ox.ac.uk, robert.simpson@astro.ox.ac.uk}

\altaffiltext{1}{Dept. of Astronomy \& Astrophysics, University of Chicago, 5640 S. Ellis Ave.,
Chicago, IL 60637, USA}

\begin{abstract} 
Yellowballs are a collection of approximately 900 compact, infrared
sources identified and named by volunteers participating in the Milky
Way Project (MWP), a citizen-science project that uses GLIMPSE/MIPSGAL
images from \emph{Spitzer} to explore topics related to Galactic star
formation. In this paper, through a combination of catalog
cross-matching and infrared color analysis, we show that yellowballs
are a mix of compact star-forming regions, including ultra-compact and
compact \ion{H}{2} regions, as well as analogous regions for less
massive B-type stars. The resulting MWP yellowball catalog provides a
useful complement to the Red \emph{MSX} Source (RMS) survey. It
similarly highlights regions of massive star formation, but the
selection of objects purely on the basis of their infrared morphology
and color in \emph{Spitzer} images identifies a signature of compact
star-forming regions shared across a broad range of luminosities, and
by inference, masses. We discuss the origin of their striking
mid-infrared appearance, and suggest that future studies of the
yellowball sample will improve our understanding of how massive and
intermediate-mass star-forming regions transition from compact to more
extended bubble-like structures. 
\end{abstract}


\section{INTRODUCTION} \label{sec:intro}

The Milky Way Project (MWP; \citealt{sim12}) is one of a suite of
highly productive online citizen science initiatives in the
Zooniverse, developed and maintained by the Citizen Science Alliance
\citep[e.g.][]{lin08, smi11, for12}. The first implementation of the
MWP utilized archived \emph{Spitzer} GLIMPSE/MIPSGAL images
\citep{ben03, miz08, chu09, car09} to study star formation over one
third of the Galactic plane via the categorization of infrared
``bubbles'', which are characteristic of \ion{H}{2} regions and their
associated photodissociation regions (PDRs). The PDRs are prominent in
the IRAC  8~$\mu$m band, which traces emission from polycyclic
aromatic hydrocarbons (PAHs), while MIPS images of bubble interiors
often show 24~$\mu$m emission, which is likely associated with thermal
emission from dust grains within \ion{H}{2} regions
\citep[e.g.,][]{wat09}. The principal task of citizen science
volunteers entailed using an ellipse-drawing tool to mark the sizes,
orientations, ellipticities, and thicknesses of \ion{H}{2} region/PDR
features. The first data release expanded the previous bubble catalogs
of \citet{chu06, chu07} by nearly an order of magnitude
\citep{sim12}. Furthermore MWP citizen science classifications have
been used as training sets for a machine learning algorithm \citep{bea14}.

Demonstrating the serendipitous nature of citizen science efforts,
volunteers went beyond their assigned tasks and started tagging and
discussing, using the MWP `Talk' interface, compact yellow objects
(``yellowballs'') in the GLIMPSE/MIPSGAL images shortly after the MWP
opened to the public. The term is descriptive of their color and
compact appearance in the GLIMPSE/MIPSGAL images, which use a
4.5~$\mu$m (blue), 8~$\mu$m (green), and 24~$\mu$m (red)
representative color scheme\footnote{All MWP images viewed by
  volunteers are available at http://mwp-milkman.herokuapp.com/}. In
total 928 yellowballs were identified by MWP participants (see
Table~\ref{tbl:ybo}). Most yellowballs appear in three types of
environments: (1) as isolated objects in filamentary infrared dark
clouds (IRDCs), along with bright 24~$\mu$m point sources that are
typically associated with embedded massive protostars
\citep[e.g.,][]{rat10, bat14}; (2) clustered at the intersection of
bipolar bubbles, which have been associated with outflows from massive
protoclusters, and are often perpendicular to filamentary IRDCs; and
(3) in bubble hierarchies, often along the rims of large bubbles, many
of which have been associated with known \ion{H}{2} regions. 

In Sections~\ref{sec:ybo} and \ref{sec:photo} we use the spatial
distribution of yellowballs, cross-matches with existing catalogs of
star-formation tracers, and mid- and far-infrared photometry to show that the
yellowballs identified in the MWP are a collection of objects tracing
a compact, dense phase of massive (O- and B-type) star
formation. This includes a mix of compact and ultra-compact \ion{H}{2}
regions as well as analogous regions for less massive B stars. We then
examine in Section~\ref{sec:discuss} the origin of the striking
mid-infrared appearance of yellowballs and discuss how they fit into
our current picture of massive star formation. Finally, a summary of
our findings and our conclusions are presented in
Section~\ref{sec:conc}.

\section{THE YELLOWBALL -- STAR-FORMING REGION CONNECTION}\label{sec:ybo}

\subsection{The Galactic Distribution of Yellowballs}

\begin{figure}
\plotone{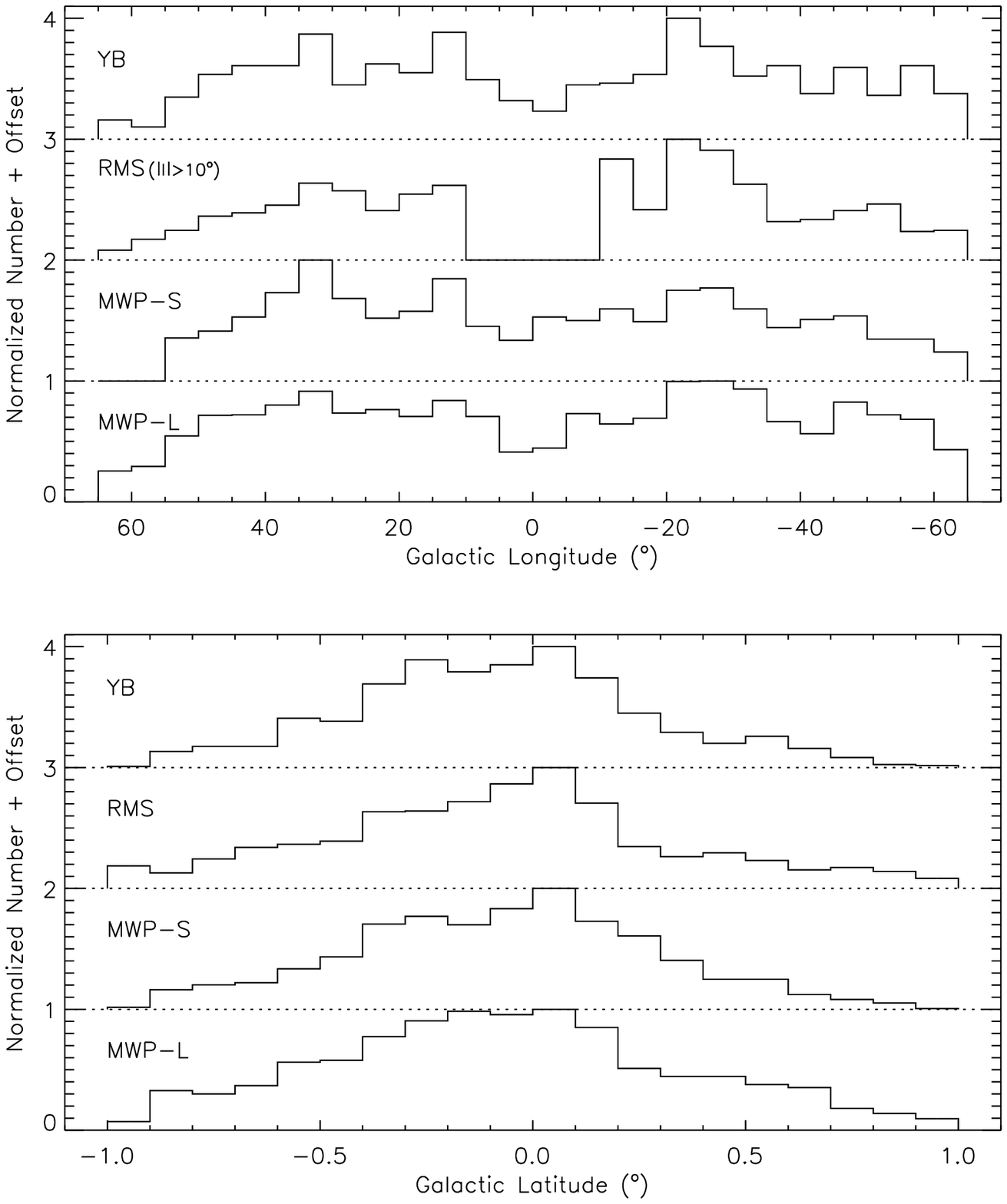}
\caption{Galactic longitude (upper) and latitude (lower) distribution
  of yellowballs (YB). The distributions are similar to the other
  tracers of Galactic star formation shown: young RMS objects (RMS), MWP large bubbles (MWP-L), and MWP
  small bubbles (MWP-S). Note RMS does not cover $|l|<10\degr$. Each
  histogram has been normalized to its largest bin value and an
  offset of 0, 1, 2, and 3 has been added to the MWP-L, MWP-S, RMS and
  YB histograms respectively.
  \label{fig:lb}}
\end{figure}

Figure~\ref{fig:lb} shows the Galactic longitude and latitude
distribution of yellowballs, along with the distribution of MWP
bubbles, and young Red \emph{MSX} Source (RMS) objects, which have
been shown to be tracers of Galactic star formation activity
\citep{ken12,sim12,lum13,urq14}.

Each histogram was normalized to its largest bin value and offsets
were applied in order to facilitate comparison. It is readily apparent
that the yellowball distributions have many similarities with the MWP
bubble and RMS object distributions. In particular, the mean of the
distribution in Galactic latitude is slightly below the Galactic
mid-plane ($-0\fdg08$), and it also peaks slightly above the Galactic
mid-plane (in the $0\fdg0$ -- $0\fdg1$ bin).  In longitude we see
common features associated with Galactic structure, such as peaks
around $l=+30\degr$ -- $+40\degr$ and $l=-20\degr$ -- $-30\degr$, as
well as a steady decrease in the number of sources beyond
$l\sim+40\degr$.  In short, the Galactic longitude and latitude
distribution of yellowballs supports them being a population of
sources associated with Galactic star formation.

\subsection{Association of Yellowballs with Dense Clumps/Cores}

We used the Virtual Astronomical Observatory (VAO) application {\sc topcat}
\citep{tay05} to cross-match the list of yellowball positions with the
ATLASGAL catalog of compact 870~$\mu$m sources \citep[average radius
$14\arcsec$, $\sigma=3\arcsec$;][]{cse14} and the catalog of
1.1~mm sources (average radius $51\arcsec$, $\sigma=24\arcsec$) from
the second data release of the Bolocam Galactic Plane Survey
\citep[BGPS;][]{agu11, dun11, gin13}. Using a match distance
comparable to the average size (diameter) of yellowballs ($0\farcm4$,
$\sigma \sim 0\farcm3$) we find that 245 (49\%) of the 502 yellowballs found
within the BGPS survey region have a BGPS association. The ATLASGAL survey
encompasses the entire MWP region, and, using the same match distance,
we find 524 (56\%)  of the 928 yellowballs have matches with an ATLASGAL compact
source. The catalog identifications for associated BGPS and ATLASGAL sources are listed in Columns~2
and 3 of Table~\ref{tbl:ybocat}.

For comparison, cross-matches done between 20 randomly
generated datasets (with statistically the same latitude and longitude
distribution as the yellowball sample) had only $6\pm2$ ($1\pm0.4$\%) 
and $10\pm2$ ($1\pm0.2$\%) BGPS and ATLASGAL associations on average
respectively.  We conclude that any contamination from random
associations is minimal, and the proximity of yellowballs to
regions of dense molecular gas is again consistent with what would be
expected for a population of objects associated with star formation activity.

\subsection{Association of Yellowballs with \ion{H}{2} Regions}

We performed a similar analysis to cross-match the yellowball positions with the \emph{WISE}
catalog of Galactic \ion{H}{2} regions \citep{and14}. We found that
599 (65\%) yellowballs have matches to within $0\farcm4$ (see Column~5
of Table~\ref{tbl:ybocat}). Entries in
the \emph{WISE} \ion{H}{2} region catalog were selected by employing
mid-infrared criteria similar to those used to identify ``bubbles'' in
the MWP and \citet{chu06,chu07} catalogs hence it is likely not
complete for very compact sources. Their resulting catalog
differentiates between four classes of objects (K, G, C, Q): known
\ion{H}{2} regions, which have measured Radio Recombination Line (RRL)
or H$\alpha$ spectroscopic emission (K); grouped \ion{H}{2} regions,
where candidates are associated via positional correlation with known
multiple \ion{H}{2} regions (G); candidate \ion{H}{2} regions, which
have characteristic \ion{H}{2} region mid-infrared morphology
spatially coincident with detected radio continuum emission but lack
RRL or H$\alpha$ observations (C; follow-up observations by Anderson
and colleagues suggest that essentially all objects in group C are
bona fide \ion{H}{2} regions); and radio quiet objects, which may
contain only intermediate-mass stars, or be \ion{H}{2} regions in
either early or late stages of evolution (Q). However, since most
of the radio quiet objects have small angular sizes and correlate with
cold dust, they are probably in the earliest phases of \ion{H}{2}
region evolution \citep{and14}. Yellowballs with \emph{WISE}
\ion{H}{2} region catalog matches span all four categories, with
associations as follows: 185 K (31\%), 62 G (10\%), 144 C (24\%), and
208 Q (35\%). The \emph{WISE} source class for each matched yellowball
is shown in Column~6 of Table~\ref{tbl:ybocat}. 
 
\subsection{Association of Yellowballs with Red \emph{MSX} Sources}

The RMS catalog is the largest statistically selected
catalog of young massive protostars and \ion{H}{2} regions to date
\citep{lum13}. It correlates spectral information across a wide range
of wavelengths from the near-infrared to the radio regimes, uses
rigorous color criteria to classify different types of objects, and,
where possible, includes kinematic distance estimates and bolometric
luminosities for catalog entries. The final catalog lists 11
categories of objects, including five categories associated with
evolved star groups (generic evolved stars, planetary nebulae (PNe), proto-planetary
nebulae, OH/IR stars, and carbon stars); four categories of young
objects (YSOs, \ion{H}{2} regions, \ion{H}{2}/YSO, and diffuse
\ion{H}{2} regions); and two categories of ambiguous objects
(young/old sources and other). The catalog is thought to be complete
for the detection of a B0~V star at the distance of the Galactic center,
although inclusion in the catalog is dependent upon detection of the
source by \emph{MSX} and specific RMS color criteria (such as rising flux
toward longer wavelengths in the \emph{MSX} bands).

\citet{lum13} estimate that 95\% of the Galactic ultra-compact \ion{H}{2} regions
were detected, but more than 50\% of the larger compact \ion{H}{2}
regions may be missing, although some of these might have been
classified as ``diffuse \ion{H}{2} regions'' in the RMS
catalog. Therefore, the RMS classification ``\ion{H}{2} region''
should generally be interpreted as ``ultra-compact \ion{H}{2} region'', while
``diffuse \ion{H}{2} region'' may in fact include compact \ion{H}{2}
regions.

\begin{figure}
\plotone{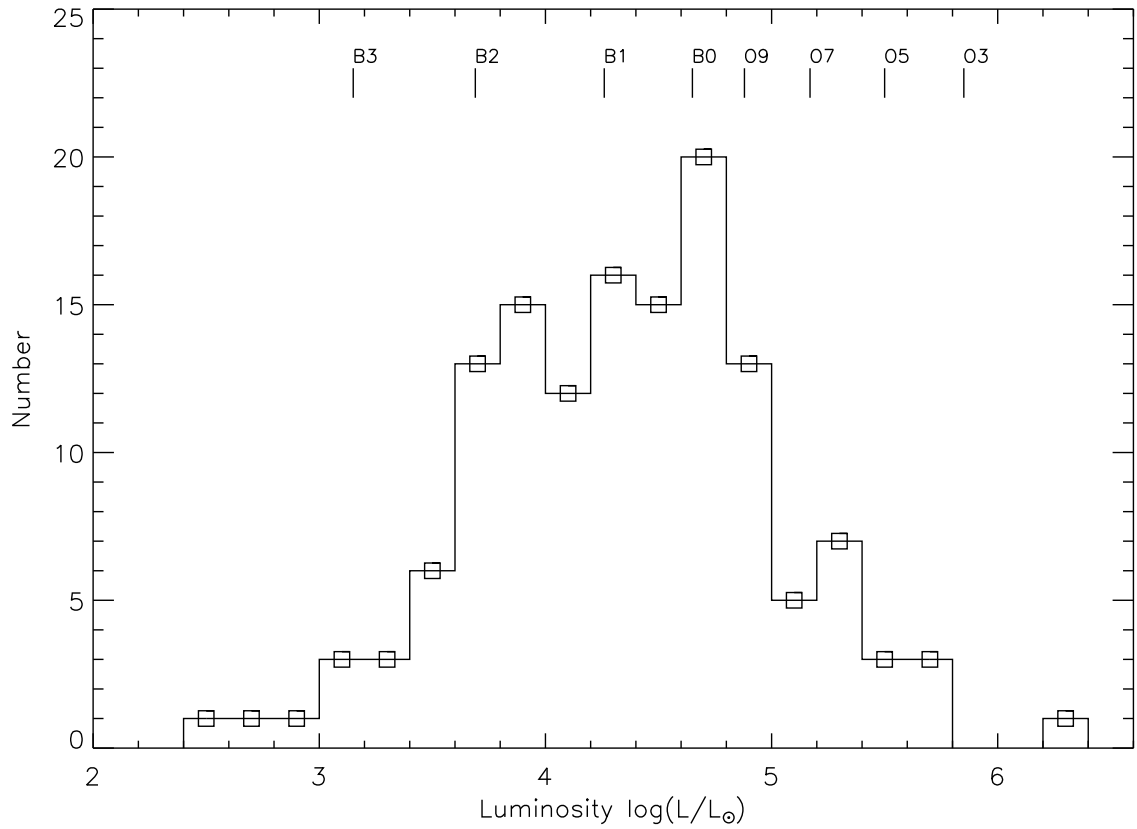}
\caption{Yellowball Luminosity. Luminosities from \citet{lum13} are
  shown for a subset of 138 yellowballs with RMS associations (coincident to within
  5$\arcsec$). For reference the ZAMS luminosity calibration of
  \citet{cro05} is shown along the upper axis. \label{fig:lum}}
\end{figure}

To investigate association of yellowballs with RMS objects, we first
cross-matched the yellowball list with the RMS catalog using a
cross-match distance corresponding to the size of each individual
yellowball (Urquhart, private communication). Of the 825 yellowballs
that overlap the RMS catalog (103 yellowballs lie within 10$\degr$ of
the Galactic center, and would therefore not be included in the RMS
catalog), 282 ($\sim$ 34\%) have RMS matches (see column 4 of Table~\ref{tbl:ybocat}). Of the 282 yellowballs
with RMS matches, 155 (19\%) are positionally coincident with RMS
objects to within 5$\arcsec$, indicating the yellowball is the main
contributor to the luminosity of these sources. Of these 155
yellowballs, 18 are classified as ``Diffuse HII region'', 4 as
``HII/YSO'', 17 as ``YSO'', and the remaining 116 as ``HII region'' in
the RMS catalog. No yellowballs are associated with any of the five
evolved star groups or ambiguous objects in the RMS catalog. All but 17 of
the 155 yellowballs have bolometric luminosity estimates; these
span a broad range from 3.30$\times 10^2$ - 1.8$\times 10^6$ L$_{\odot}$ (see
Figure~\ref{fig:lum}), with the vast majority of the yellowballs
having luminosities expected for regions of massive star formation.

\begin{figure}
\plotone{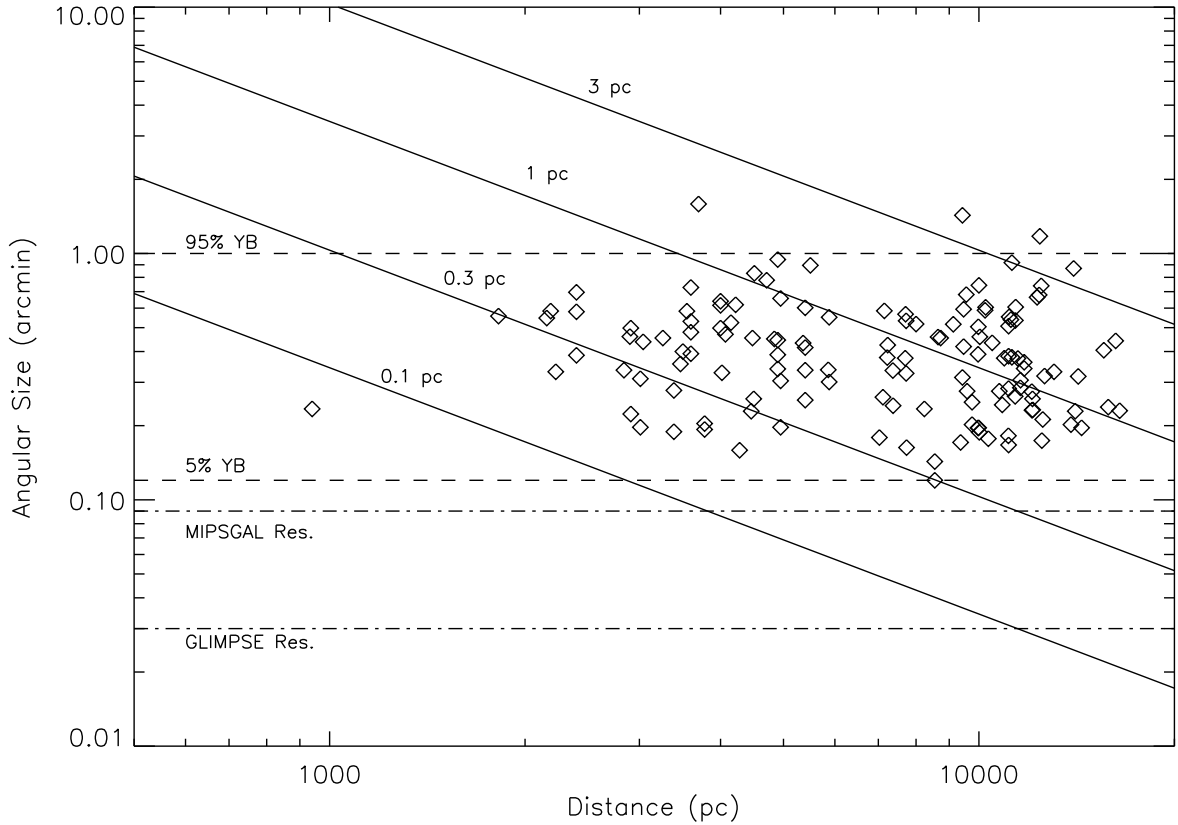}
\caption{Yellowball Angular Size and Distance. Points are plotted for the
  138 yellowballs (YB) coincident to within $5\arcsec$ with RMS sources. The horizontal dashed lines show the
  $5^{\rm{th}}$ and $95^{\rm{th}}$ percentiles for the entire
  yellowball sample (e.g., 95\% of all yellowballs have an angular
  size $\lesssim 1\arcmin$.   The diagonal solid lines represent
  objects with a constant physical size (as labeled). For
  reference the nominal MIPSGAL and GLIMPSE resolutions are shown as
  horizontal dash-dot lines.
\label{fig:asize}}
\end{figure}

Figure~\ref{fig:asize} shows the distribution of
physical sizes for the same sample of 138 yellowballs. For each
yellowball an effective angular size was calculated using the average
of $\Delta l$ and $\Delta b$ from Table~\ref{tbl:ybo} then converted
to a physical size using distances from the RMS catalog. It should be
kept in mind that these sizes are strict upper limits to the sizes of
any ionized regions within these objects for two reasons: (1) the size
of each yellowball corresponds to the size of the user-drawn rectangle
enclosing the yellowball, and (2) the yellowball is tracing the maximum extent of
the PDR associated with each source so any enclosed \ion{H}{2} region
is necessarily smaller (see Section~\ref{sec:discuss}). This suggests that many of the smaller
yellowballs are likely to contain ultra-compact \ion{H}{2} regions or perhaps
even hypercompact \ion{H}{2} regions, identified with the earliest
manifestations of ionized gas around young massive stars
\citep{fra00}. 

\begin{figure}
\epsscale{.62}
\plotone{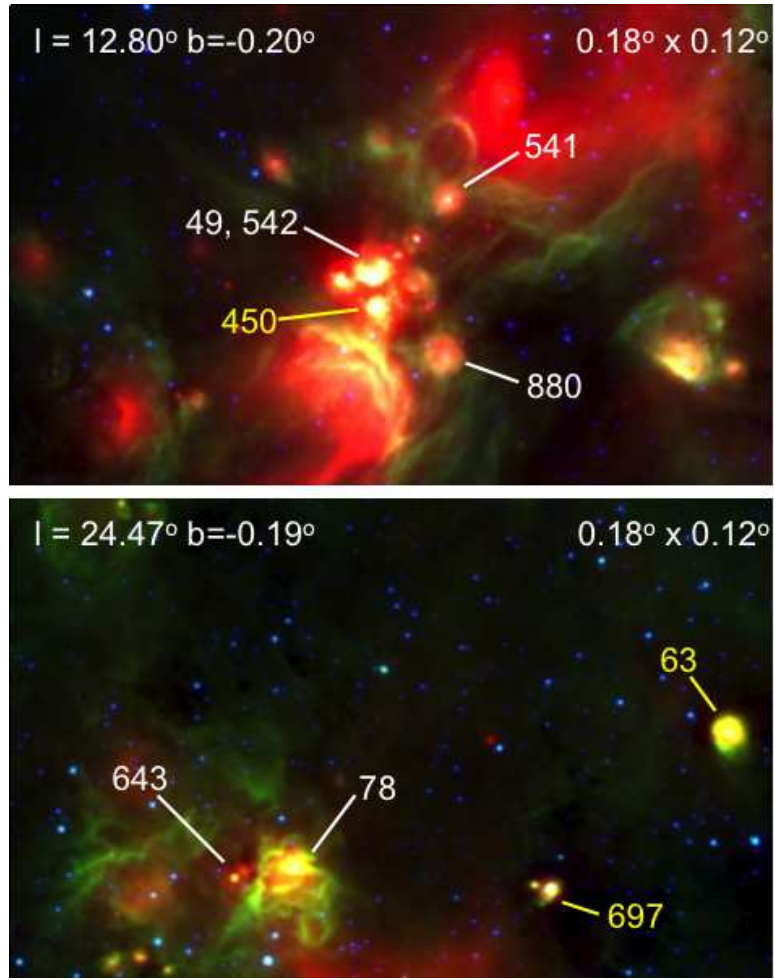}
\caption{MWP GLIMPSE/MIPSGAL images, 4.5~$\mu$m (blue), 8~$\mu$m
  (green), and 24~$\mu$m (red), showing yellowballs with (in white)
  and without (in yellow) RMS counterparts. The top panel shows part
  of W~33, with a cluster of yellowballs located between two large
  bubbles. The lower panel shows two isolated yellowballs and two
  yellowballs associated with a larger \ion{H}{2} region (see the main
  text for details). The Galactic coordinates of the image center and
  the angular size of the image are shown at the top of each frame.  \label{fig:ybo}} 
\end{figure}

Figure~\ref{fig:ybo} presents two images from the MWP containing
yellowballs with and without RMS counterparts (coincident to within
5$\arcsec$). The top panel shows the central portion of W~33, with a
cluster of yellowballs located between two large infrared
bubbles. Yellowballs 541 and 880 are classified as ``YSO'' and
``diffuse \ion{H}{2} region'', respectively, in the RMS catalog, while
yellowballs 49 and 542 are classified as \ion{H}{2} regions. The latter are
unresolved near the center of the image, with a ``lima-bean'' morphology.
Yellowball 450 is a known \ion{H}{2} region, first detected as a 20
$\micron$ source by \citet{dyc77} and characterized at cm radio
wavelengths by \citet{has83}, but it was not included in the RMS
catalog because it was not included in the \emph{MSX} catalog.

In the lower panel,  yellowballs 78 and 643 both have RMS
counterparts, and are classified as an ``\ion{H}{2}
region'' and ``YSO'' respectively. In contrast, the other
yellowballs in the field (63 and 697) fail the color
criteria for inclusion in the RMS catalog: both have  \emph{MSX} F(8
$\micron$) $>$ F(14 $\micron$), which is likely due to them having a  large PAH ionization
fraction (see Section~\ref{sec:discuss} for more discussion). 

The MWP yellowball catalog provides a useful
compliment to the RMS survey. It similarly highlights regions of
massive star formation, but our results suggest that the selection of
objects purely on the basis of their infrared morphology and color in
the higher-resolution \emph{Spitzer} images (cf. \emph{MSX})
identifies a signature of compact star-forming regions shared across a
broad range of luminosities, and by inference, masses.

\section{PHOTOMETRY} \label{sec:photo}

We performed an infrared color analysis in order to explore further
the association of yellowballs with star-formation activity apparent
both by visual inspection of MWP images and by the numerous catalog
cross-matches outlined in the previous section. To rule out the
possibility that these compact objects might be PNe, we used color
criteria developed by \citet{and12}. In particular, \citet{and12}
showed that the 12~$\mu$m to 8~$\mu$m flux ratio, derived from
\emph{WISE} and \emph{Spitzer} IRAC data respectively, is a robust
discriminator for separating \ion{H}{2} regions and planetary nebulae
(PNe). The IRAC 8~$\mu$m band is very sensitive to emission associated
with the PAH bands centered at 7.7 and 8.6~$\mu$m. In contrast,
although the WISE 12~$\mu$m band does cover the 7.7, 8.6, 11.3 and
12.7~$\mu$m PAH bands, its normalized response is relatively low
($\sim 0.4 - 0.8$) at these wavelengths, and the peak sensitivity of
the band lies at 14.5~$\mu$m, longwards of most of the PAH
emission. Given this combination of PAH and filter properties,
\ion{H}{2} regions tend to have a lower 12/8 flux ratio than PNe
because their mid-infrared emission is dominated by strong, broad PAH
features, whereas PAH features in PNe tend to be narrower and
relatively weaker \citep{bre89}.

\subsection{Selection Criteria \& Technique}
We performed aperture photometry on a representative sample of 183 yellowballs using
\emph{Spitzer} 8 $\mu$m \& 24 $\mu$m and \emph{WISE} 12 $\mu$m band
data. This sample was chosen using all yellowballs coincident with RMS sources
(to within 5$\arcsec$) that were not saturated in MIPSGAL images at 24 $\mu$m (N=81),
and a comparable number of yellowballs without RMS matches
(N=102). The latter sample was chosen by selecting yellowballs with
the highest (N=52) and lowest (N=50) hit rates, where the hit rate is
defined by the ratio of the number of times a yellowball was
identified by a volunteer to the number of times it was viewed.
The rationale for choosing this sample was to allow us to investigate any  potential
differences based on hit rate.  Only objects viewed $>50$ times, which
have hit rates $>0.1$, are included in the list of 928 yellowballs. 

Aperture photometry was done using the IDL-based {\sc imview} program \citep{hig97}. For
extended infrared sources at low Galactic latitude, especially those
associated with star-formation activity, background estimation and
subtraction is the most important source of photometric error as the
infrared emission in the Galactic plane is highly structured
\citep[e.g.,][]{fic96}.  {\sc imview} allows the user to select points
that define the shape of the photometric aperture, and that constrain
the surface fit used to estimate the background. The background fit
can then be varied by selecting different background points and/or
using different interpolating functions. For each source the average
flux, using two different apertures and two different surface fits,
was obtained and the standard deviation of the measurements was used
as a measure of the uncertainty.  

\subsection{Results}

The average flux density and uncertainty at 8, 12, and 24~$\mu$m  for each
yellowball is reported in columns 2 -- 7 of
Table~\ref{tbl:ybophot}. The table is divided into three subsections
corresponding to the RMS-match sample, and the high and low hit rate
non-RMS samples. In addition,  Columns 8 -- 11 of
Table~\ref{tbl:ybophot} indicate whether the source has a cross-match
in another catalog, and gives the \emph{WISE} catalog class if applicable.

\begin{figure}
\epsscale{1.0}
\plotone{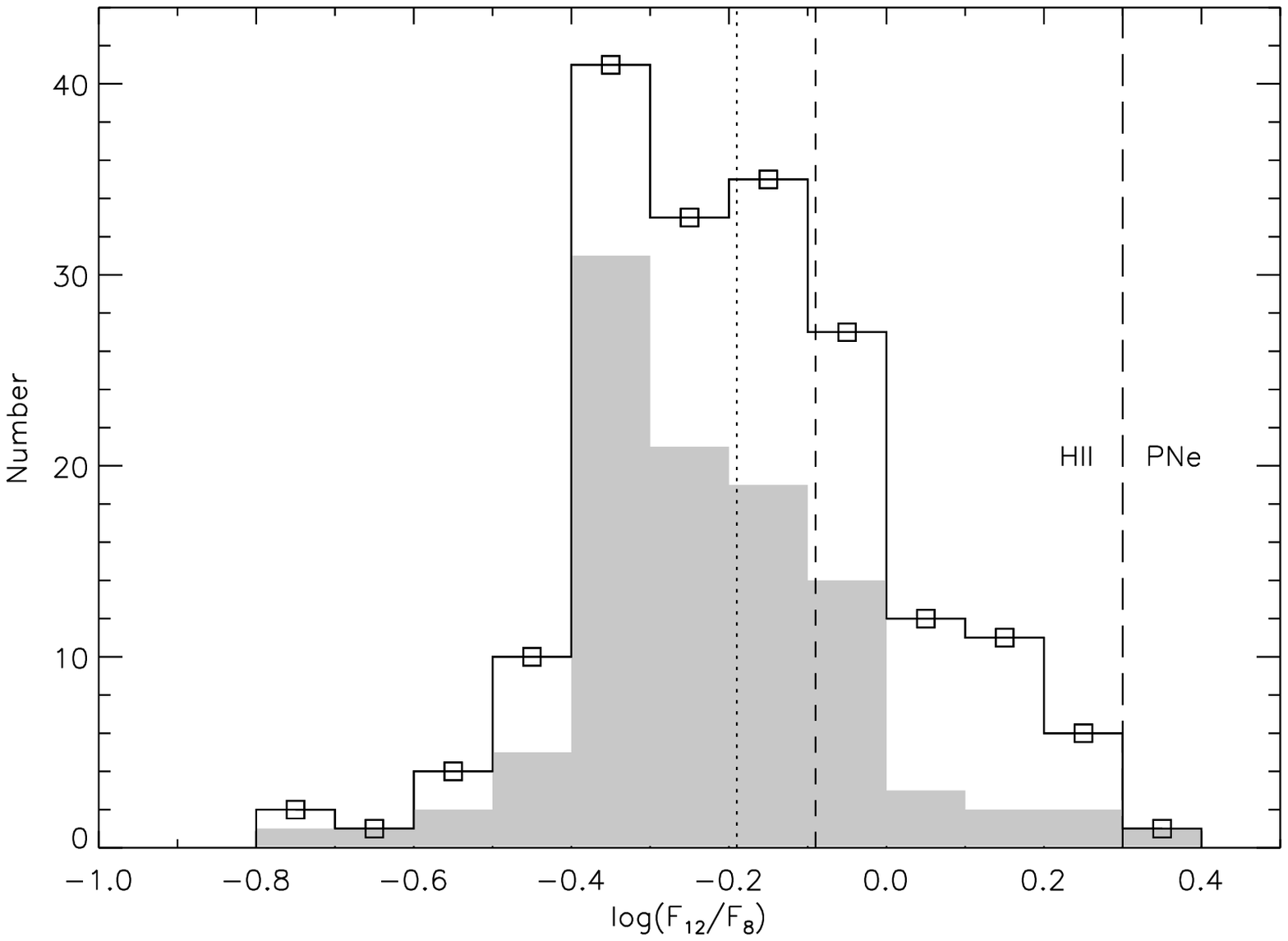}
\caption{\emph{WISE} and \emph{Spitzer} photometry for a sample of 183
 yellowballs (see the main text for selection criteria). The dotted
 line shows the average color ($-0.19$), which can be compared with
 the average color of the sample of \ion{H}{2} regions from \citet{and12} shown as
  a short-dash line ($-0.09$). The long-dash line shows the \ion{H}{2}
  region -- PNe cutoff from \citet{and12} at $+0.3$. The grayed
  region shows the distribution of yellowballs without RMS
  associations (N=102; average color = $-0.23$). \label{fig:wise}}
\end{figure}

In Figure~\ref{fig:wise} we show a histogram of the resulting
$\log(F_{12}/F_{8})$. The average value for this sample ($-0.19$), and
the average color for the sample of \emph{WISE} \ion{H}{2} regions
($-0.09$) from \citet{and12} are indicated. The photometric
uncertainties are typically about half of a bin width ($\sim0.05$),
and the entire sample (except for one yellowball with amorphous
boundaries) satisfies the robust F(12 $\mu$m)/F(8 $\mu$m) flux ratio
criterion determined by \citet{and12} to separate \ion{H}{2} regions
from PNe at $+0.3$. We note that the RMS and non-RMS subsamples do not
have identical distributions, specifically the average color of the
non-RMS yellowballs ($-0.23$) is more negative than the average of the
RMS-matched subsample ($-0.13$), and a KS-test shows that the
two samples are significantly different (significance level
$p=0.001$). We also find that, within the non-RMS sample, the high hit
rate sample has an average color ($-0.28$) that is more negative than
the low hit rate sample ($-0.17$). Again, a KS-test shows the two
samples are signficantly different (significance level $p=0.001$).  We
discuss the likely cause of these color differences in the next section.

We found that yellowballs have on average $F_{24} \sim 3.5
F_{8}$ ($\sigma = 2.5$, median  $F_{24}  \sim 2.9 F_{8}$). This is consistent
with the average \ion{H}{2} region spectrum shown in \citet{and12}.

\section{DISCUSSION}\label{sec:discuss}

The distinct mid-infrared appearance of the yellowballs in the
GLIMPSE/MIPSGAL images used in the MWP is not primarily due to the
rough equality of the 8~$\mu$m and 24~$\mu$m fluxes mentioned in
Section~\ref{sec:photo}, as this is expected for all \ion{H}{2}
regions, but comes about because the emission is spatially
coincident. This spatial coincidence is expected in the early stages
of the evolution of \ion{H}{2} regions/PDRs.  Models predict that large PDRs will form around any initial
(dust-filled) ionized region, and that the maximum size of the
PDR, and the time at which the maximum size is obtained, are both
relatively insensitive to stellar luminosity \citep{rog92}. For
example, Figure 9 of \citet{rog92} shows that the PDR associated with
an \ion{H}{2} region expanding into a molecular cloud having densities
between 300 to  3000 cm$^{-3}$ will reach a maximum thickness of order 0.1 to 1 pc on timescales of
$10^4 - 10^5$ years for a wide range of stellar
luminosities. This size range of the model PDR, and the insensitivity
of the model to stellar luminosity, are both consistent with the derived yellowball
sizes and the range of luminosities presented in Section 2.4.

As the region evolves the ionized region is expected to catch-up to
the photo-dissociation front resulting in a thin, shocked \ion{H}{1}
region/PDR surrounding the \ion{H}{2} region.  Much of the dust will
be removed from the central portion of the \ion{H}{2} region via the
action of radiation pressure and stellar winds \citep{dra11}, and PAHs
will be destroyed within the ionized gas \citep{gia94}. This leads to
a clear spatial separation between the $F_{8}$ emitting region
(PAH-rich, PDR) shown as green in the MWP images, and the $F_{24}$
emitting region (depleted interior hot dust, perhaps resupplied by the
erosion of denser clumps in the region; \citealt{eve10}) shown as red
in the MWP images.

We noted in Section~\ref{sec:photo} that the yellowball sample has an
average $\log(F_{12}/F_{8})$ color that is more negative than the
average color for the general \ion{H}{2} region sample. This is most
likely due to the fact the yellowball sample contains a higher
fraction of compact objects with a large PAH ionization fraction
(Ybarra 2014, private communication). \citet{roe96} showed that
compact \ion{H}{2} regions had a much higher 7.6/11.2 PAH intensity
ratio (ranging from 5 -- 11) compared to more evolved \ion{H}{2}
regions ($\sim 3$). This effect is related to the PAH ionization
fraction because the strength of the PAH bands around 7-8~$\mu$m is
highly sensitive to the ionization state of the PAHs (becoming much
stronger in ionized PAHs), whereas the strength of PAH bands at longer
wavelengths are minimally affected by the ionization state
\citep{dra07}. Thus we would expect the $\log(F_{12}/F_{8})$ color to
become increasingly negative with increasing PAH ionization fraction.

This may also be the cause of the shift in the average
$\log(F_{12}/F_{8})$ color from $-0.13$ for RMS-matched yellowballs to
$-0.23$ for yellowballs with no RMS matches. The median size of the
non-RMS matched yellowballs was $0\farcm29$ compared with $0\farcm46$
for the RMS-matched sources. Assuming that the non-RMS and RMS-matched
yellowballs have a similar distribution in distance, this implies
that the non-RMS sources are more compact on average.  

Similarly we observed a difference in the average color of high
($-0.28$) and low hit rate ($-0.17$) yellowballs without RMS
matches. The median size of the high hit rate sample is $0\farcm27$
compared with $0\farcm35$ for the low hit rate sample.  For the entire
yellowball sample the correlation between hit rate and angular size is
very weak (correlation coefficient $r = -0.08$). This is not
surprising,  as there are clearly other factors such as the proximity
of the yellowball to other interesting objects, and the overall
complexity of the field, that influence hit rate.  Given this weak
correlation, if the two samples have a similar distribution in
distances, then the high hit rate sample will have a higher proportion
of physically compact objects leading to a more negative average
$\log(F_{12}/F_{8})$ color.

As touched on in Section~\ref{sec:ybo}, it is clear that the RMS
catalog does not include some yellowballs that are definitely
star-forming regions. To explore this further we cross-matched the 646
non-RMS yellowballs with the \emph{MSX} catalog using a search radius
of $10\arcsec$. Of these 199 (31\%) did not have an \emph{MSX} catalog
entry, 255 (39 \%) had poor quality (S/N $\la 5$) in band E (21.3
$\mu$m), and the remaining 192 (30 \%) failed one or more of the MSX
color cuts described in detail in \citet{lum13}. This last subset is
particularly interesting as it probably includes many compact objects
with strong PAH emission.

The catalog cross-matches presented in this paper offer opportunities
for a multitude of possible explorations. For example, consider the
138  yellowball-RMS matches that have distance and luminosity
estimates. Of these objects, 35 have BGPS counterparts and 114 have
ATLASGAL counterparts, indicating that nearly all yellowballs with RMS
entries are associated with dense gas. Objects in this group span the
range of luminosities seen in Figure 2. On the other hand, of the 208
yellowballs that are classified as radio-quiet objects in the
\emph{WISE} \ion{H}{2} region catalog of \citet{and14}, only 96 (46\%) have BGPS
or ATLASGAL counterparts and only 25 (12\%) have RMS
associations. Twenty-two of the 25 RMS-associated radio-quiet
yellowballs have luminosity estimates, all of which are $<
5\times10^{4} L_\sun$ ($\leq$ B0 ZAMS equivalent), suggesting that
many low-luminosity yellowballs were missed by these surveys. 

The most straightforward interpretation of these low-luminosity
yellowballs is that they are associated with the formation of mid- to
late- B stars, which are expected to have large, relatively long-lasting PDRs,
combined with small, weak \ion{H}{2} regions \citep{ker02,lun14}.  The
long lifetime of the PDRs in this case could explain why the
percentage of yellowballs with dense gas associations is lower than
that found for the RMS-matched sample.
Intriguingly though, it is possible that some of these sources are
massive protostars in a pre-ultra-compact \ion{H}{2} region stage; such protostars can
exhibit low overall luminosities due to low core temperatures while at
the same time having mid-infrared emission that would be visible as a
yellowball \citep[e.g., Mol 160,][]{mol08,wc12}. Further
high-resolution and high-sensitivity observations at infrared and
radio wavelengths would help to distinguish between these two options
by, for example, detecting outflows expected for massive protostars,
or by detecting the small \ion{H}{2} regions expected to be associated
with the B-type stars.

\section{SUMMARY \& CONCLUSIONS}\label{sec:conc}

We have presented multiple lines of evidence that yellowballs are a
mix of compact star-forming regions, including ultra-compact and
compact \ion{H}{2} regions, as well as analogous regions forming less
massive stars. Visual inspection of MWP images indicates these objects
are typically found in IRDCs and/or in bubble hierarchies, and their
distribution in Galactic longitude and latitude mirrors the
distribution of MWP bubbles and young RMS objects. Color analysis of
yellowballs using \emph{Spitzer} and \emph{WISE} data indicates that
yellowballs occupy regions of infrared color space that include
\ion{H}{2} regions and exclude evolved objects such as
PNe. Cross-matching the yellowballs with the ATLASGAL, BGPS,
\emph{WISE}-\ion{H}{2} region, and RMS catalogs, indicates that the
majority of these objects are unambiguously associated with dense
molecular clumps and other signposts of star formation. No yellowballs
are associated with any of the evolved star categories in the RMS catalog.

We examined the luminosity and physical size distribution of a sample of 138
yellowballs with $<5\arcsec$ positional associations with RMS catalog
sources. These objects span at least 3 orders of magnitude in
luminosity, and have luminosities consistent with massive star-forming
regions. Typical sizes for these yellowballs are comparable to compact
\ion{H}{2} regions (sub-parsec in size); however, we expect that these
sizes are upper limits on the extent of the ionized gas as the size includes
emission from the surrounding PDR. The great majority of these sources
are ultra-compact \ion{H}{2} regions or even younger/denser objects.

The origin of the distinct yellow color of these objects in the
GLIMPSE/MIPSGAL MWP images is the cospatial emission from PAHs and
dust. This is expected for the earliest stages of massive star formation when
PDRs will be at their thickest extent and various
dust-clearing/destruction mechanisms within the ionized gas have had
minimal time to act on the dust distribution. 

Yellowballs are analogous to the ``green peas'' of the archetype
Galaxy Zoo project, in the sense that they represent a class of
objects identified and recorded by citizen science
volunteers \citep{card09}. Just as the discovery and subsequent
studies of green peas have yielded critical new insights into the
evolution of galaxies \citep{amo12,cha12}, we expect that future
studies of the yellowball sample will improve our understanding of how
massive star-forming regions transition from compact embedded stages
(e.g., massive protostars and ultra-compact \ion{H}{2} regions) to more evolved
\ion{H}{2} regions. Similarly, those yellowballs associated with
slightly lower-mass star formation will provide us with a comparable
view of how these objects transition from highly embedded
stars/clusters to larger bubble structures.

%thank the referee
\acknowledgments
We thank all MWP volunteers for their enthusiastic participation in
this project, especially Sandy Harris, Ipspieler, Greg Galanos, and
Larry West. We thank Jason Ybarra for pointing out that a high PAH ionization
fraction could produce more negative $\log(F_{12}/F_{8})$ values based
on his {\sc cloudy} models, and thank Sarah Kendrew for the use of
her correlation code. We also thank James Urquhart for his assistance
with cross-matching the yellowball list and RMS catalog, and the
referee for comments that greatly improved this paper. GW-C was funded
in part through a Research Seed Grant from NASA's Illinois Space Grant
Consortium, and GW-C and KA gratefully acknowledge support from a
Brinson Foundation grant in aid of astrophysics research at the Adler
Planetarium. CK thanks Iowa State undergraduate Alicia Carter for her
assistance with this work. 

{\it Facilities:} \facility{WISE}, \facility{Spitzer}, \facility{MSX}, \facility{CSO}, \facility{APEX}

%References

%Figures

%lb histograms
\clearpage

%Luminosity  Histogram
\clearpage

%Angular Size Plot
\clearpage

% Figure Showing Select YBOs
\clearpage

% WISE/Spitzer Photometry Histogram 
\clearpage

%Tables

% Table 1 - Yellowball Basic Information 
\clearpage
\begin{deluxetable}{llccccc}
\tablewidth{0pt}
\tablecaption{Yellowballs Identified by MWP Participants \label{tbl:ybo}}
\tablehead{ \colhead{Short} & \colhead{MWP} &  \colhead{$l$} & \colhead{$b$} & \colhead{$\Delta l$\tablenotemark{a}} &  \colhead{$\Delta b$\tablenotemark{a}} & \colhead{Hit} \\
\colhead{ID} & \colhead{ID}  & \colhead{(deg)} & \colhead{(deg)} & \colhead{(arcmin)} & \colhead{(arcmin)} & \colhead{rate\tablenotemark{b}}
}
\startdata
 1 & MWP1G311565$+$00230Y & 311.565 & 0.230 & 0.160 & 0.150 & 0.78 \\
 2 & MWP1G340572$+$00360Y & 340.572 & 0.360 & 0.349 & 0.338 & 0.78 \\
 3 & MWP1G342062$+$00422Y & 342.062 & 0.422 & 0.496 & 0.543 & 0.67 \\
 4 & MWP1G032122$+$00091Y & 32.122   & 0.091 & 0.503 & 0.500 & 0.65 \\
 5 & MWP1G327901$+$00154Y & 327.901 & 0.154 & 0.259 & 0.263 & 0.65 \\
\enddata
\tablecomments{Table~\ref{tbl:ybo} is published in its entirety in the
  electronic edition. A portion is shown here for guidance regarding
  its form and content.}
\tablenotetext{a}{Average size of user-drawn rectangle enclosing the yellowball.}
\tablenotetext{b}{The ratio of number of times the yellowball was identified to number of times the yellowball was viewed.}
\end{deluxetable}

% Table 2 - Yellowball Catalog Cross-matches
\clearpage
\begin{deluxetable}{lcccccc}
\tabletypesize{\footnotesize}
\rotate
\tablewidth{0pt}
\tablecaption{Yellowball Catalog Matches and MIPSGAL 24~$\mu$m Saturation \label{tbl:ybocat}}
\tablehead{ \colhead{Short ID} & \colhead{BGPSv2\tablenotemark{a}} &
  \colhead{ATLASGAL} & \colhead{RMS} & \colhead{WISE} &  \colhead{WISE
    Class\tablenotemark{b}} & \colhead{MIPSGAL Saturated\tablenotemark{c}} 
}
\startdata
1 & \nodata & G311.5638$+$0.2298 & N & N & \nodata & N \\
2 & \nodata & G340.5742$+$0.3608 & N & G340.573$+$00.359 & Q & N \\
3 & \nodata & G342.0579$+$0.4211 & G342.0610$+$00.4200 & G342.062$+$00.417 & K & Y \\
4 & G032.119$+$00.092 & G032.1173$+$0.0909 & N & G032.123$+$00.086 & G & N \\
5 & \nodata & G327.9065$+$0.1573 & G327.9018$+$00.1538 & N & \nodata & Y
\\ 
\enddata
\tablecomments{Table~\ref{tbl:ybocat} is published in its entirety in the
  electronic edition. A portion is shown here for guidance regarding
  its form and content.}
\tablecomments{If a catalog cross-match was made the catalog
  identification is shown. N indicates no cross-match was found.}
\tablenotetext{a}{Sources not in the BGPS  survey area have \nodata
  entries.}
\tablenotetext{b}{See text for details. If no WISE cross-match was
  found \nodata is shown.}
\tablenotetext{c}{Indicates if any portion of the source was saturated
  (Y) or not (N) in MIPSGAL 24~$\mu$m images.}
\end{deluxetable}

%Table 3 - Yellowball Photometry
\clearpage
\begin{deluxetable}{lcccccccccc}
\tabletypesize{\footnotesize}
\rotate
\tablewidth{0pt}
\tablecaption{Yellowball Photometry \label{tbl:ybophot}}
\tablehead{\colhead{Short ID} 
& \colhead{$F_8$} & \colhead{$\sigma_{F_8}$} 
& \colhead{$F_{12}$} & \colhead{$\sigma_{F_{12}}$} 
& \colhead{$F_{24}$} & \colhead{$\sigma_{F_{24}}$} 
& \colhead{BGPS\tablenotemark{a}} & \colhead{ATLASGAL\tablenotemark{a}} 
& \colhead{WISE\tablenotemark{a}} & \colhead{WISE
  Class\tablenotemark{b}} \\
& \colhead{(Jy)} & \colhead{(Jy)} 
& \colhead{(Jy)} & \colhead{(Jy)} 
& \colhead{(Jy)} & \colhead{(Jy)} 
}
\tablecolumns{11}
\startdata
\cutinhead{RMS Matches}
46 & 0.44 & 0.05 & 0.31 & 0.05 & 3.23 & 0.22 & \nodata & Y & Y & C \\
65 & 2.61 & 0.04 & 2.00 & 0.05 & 6.73 & 0.17 & Y           & Y & Y & G \\
68 & 0.64 & 0.09 & 0.67 & 0.12 & 3.92 & 0.71 & \nodata & N & Y & C \\
\cutinhead{No RMS Match -- High Hit Rate}
1 & 0.05 & 0.01 & 0.03 & 0.01 & 0.81 & 0.02 & \nodata & Y & N & \nodata \\
2 & 1.32 & 0.05 & 0.83 & 0.08 & 1.24 & 0.04 & \nodata & Y & Y & Q \\
4 & 5.35 & 0.22 & 3.75 & 0.24 & 15.04 & 0.53 & Y & Y & Y & G \\
\cutinhead{No RMS Match -- Low Hit Rate}
847 & 0.023 & 0.002 & 0.011 & 0.003 & 0.011 & 0.002 & \nodata & N & N & \nodata \\
848 & 0.074 & 0.005 & 0.07 & 0.01 & 0.29 & 0.03 & \nodata & N & N & \nodata \\
850 & 0.82 & 0.05 & 0.67 & 0.04 & 1.76 & 0.07 & \nodata & N & Y & Q \\ 
\enddata
\tablecomments{Table~\ref{tbl:ybophot} is published in its entirety in the
  electronic edition. A portion is shown here for guidance regarding
  its form and content.}
\tablenotetext{a}{Y=Catalog match, N= No catalog match, \nodata =
  source not in survey area. See Table~\ref{tbl:ybocat} for match
  source ID}
\tablenotetext{b}{\nodata is shown for sources without a WISE match.}
\end{deluxetable}

\end{document}